# Which is the ontology of Dark Matter and Dark Energy?

Jorge E. Horvath
foton@iag.usp.br
Universidade de São Paulo USP - IAG Astronomy, Brazil

We adopt in this work the idea that the building blocks of the visible Universe belong to a class of the irreducible representations of the Poincare group of transformations (the "things") endowed with classificatory quantum numbers ("the properties"). After a discussion of this fundamentality, the question of the nature of both "dark" components of the Universe which are deemed necessary, but have not been observed, is analyzed within this context. We broadly discuss the ontology of dark matter/dark energy in relation to the irreducible representations of the Poincare group + quantum numbers, pointing out some cases in which the candidates can be associated to them, and others for which a reclassification of both the dark and visible (ordinary) components would be needed.

**1. Introduction.**

In spite of the enormous progress achieved in the last decades in the study of the Universe, its early stages and evolution as a physical system, an important fact revealed from a class of modern observations brings a quandary about its composition, quite reminiscent of a problem of the ancient Greek thought. At ancient times, even with available visions opposed to the elements doctrine (notably the Greek atomism, see Stanford 2020a), several centuries had to pass until the latter view was overseded. Eventually the "modern" atomism (Satnford 2020d) emerged and  supported by Dalton and co-workers from an empirical point of view, accepted as a compelling explanation of the constituent of the physical world, certainly with a debt to the ancient Greek "philosophical" atomism of Democritus and Leucippus, Newton and others, but then firmly rooted on empirical observed facts after a relatively short dispute and strong defense from Daltonian chemists. The difference between modern atomism and substance theory has been stressed in the literature (Stanford 2020b). Roughly speaking, and without a finer analysis, we may characterize substance theory as a macroscopic feature, and atomism (in its present "modern" form, quanta interacting in a vacuum state) as quite opposite to the former, since the emerging properties of matter are thought to arise from *interactions*, first among quanta and then among the sets of quanta termed "atoms" and "molecules".

The point-like structure of known particles (electrons and so on) probed up to scales ~ $10^{-5}$ (www.frontiersin.org/articles/10.3389/fphy.2020.00213/full), much smaller than their classical sizes qualifies as a provisional proof of their character of "mereological atoms" in the philosophical sense. However, the search for substructure continues and, quite amazingly, a "top down" approach (M-theory and variants) emerged to understand the visible world as vibration modes of mathematical entities (strings), with a strong neo-Pythagorean flavor. But with the evidence and elements at hand, it is valid to ask whether a metaphysical foundationalism (Stanford 2020c and references therein) can be constructed, which are the elements of its ontological categorization, and whether the rest of the unknown world (dark matter and dark energy, see below) exists and how does it fit into it. We offer a general view of these issues in this paper.

The other important aspect of this big puzzle is, of course, the view of the Universe itself, its origin and evolution. The contributions of many philosophers, thinkers and scientists along human history are too long to be summarized (see Campion 2015), but it is widely recognized to have started in its present form after Newton and contemporaries. Two centuries of "modern" (i.e. Newtonian and post-Newtonian) cosmology produced a mathematical picture of the Universe based on many assumptions, giving rise to the present Friedmann-Robertson-Walker homogeneous and isotropic models based on General Relativity around a century ago. The 20th century produced also a large deal of technological and conceptual advances that allowed, for the first time, to peep into the "hidden" assumptions of cosmology by measuring an increasing variety of physical observables (galaxies, large-scale structure, background radiation, etc.) that shifted cosmology from *mathematical* to *physical*, including the formulation of the evolution from an initial hot, dense state termed the *Big Bang* by F. Hoyle (1949), a name given with a great deal of irony at that time. Finally, the present (late) Universe is much more amenable to direct studies, and this is why the unfolding of the recent structures and related "cold" cosmic era physics are presently known in much more detail.

As an important benchmark for our discussion, a major breakthrough in Cosmology as a whole happened in 1998, when two independent groups studying distant supernovae (exploding stars undergoing thermonuclear carbon ignition) as "standard candles" which trace the expansion of the Universe, announced that their data contained evidence that the expansion rate *accelerated* lately (Riess *et al*. 1998, Perlmutter & Schmidt 2003). Within the widely accepted mathematical framework based on the General Relativity equations, the expansion needs quite particular conditions for this to happen. And this is where the question of the matter/energy content gains an additional twist, since not only the "old", early-20th century hypothesis of the existence of a *dark matter*, somewhat concentrated into structures stands today, but also a new, even stranger (*dark energy*) component widespread along the whole Universe volume was postulated to explain this observed accelerated expansion as the most simple solution. The purpose of this article is to discuss the ontological status of both components, in comparison with the "ordinary" matter mentioned above, and also to point out that there are hints of a deeper revision of the whole *classification* of the physical content of the Universe, to which dark matter/dark energy may belong. The rest of the present work is devoted to sketch some features of this attempt to understand the content of the physical world according to scientific ideas (Zinkernagel 2002).

**2. What are the Dark Matter and the Dark Energy (Phenomenology)?**

The idea that the Universe contains matter that does not "shine" (i.e. does not emit or absorb photons) goes back to a set of observations by F. Zwicky in the 1930 decade. Zwicky (1933) was studying the Coma cluster of galaxies and observed that the visual estimation of the total matter within them was not enough to produce a gravitational force strong enough for the system to be bound. Thus, he conjectured that the total amount of matter was actually *much larger*, but most of this "extra" component was unseen by the telescopes. Hence the name "dark matter" (although sometimes the equivalent expression "missing light" is used), a puzzle now more than 80 years old. Of course, if this is true, there is a huge need to understand which is its nature. In the following years, and even in contemporary observations, the presence of dark matter has been inferred in a variety of ways, and more importantly, its abundance must be *much larger* than ordinary matter that does shine by interacting with photons, while its exact fraction depends on the considered system (sometimes outnumbering the ordinary matter by about 100 parts to 1). This problem could not be solved by three or four generations of astronomers and physicists, in the sense that all kind of techniques were attempted to detect the dark component, from direct scattering experiments of ultra-low noise (where a dark matter particle is expected to collide with ordinary targets) to searches in all the electromagnetic spectrum; and theoretical constructions ranging from radical ideas to minimal extensions of the Standard Model of particle physics put forward to guide the former efforts. In spite of some recently promising claims (Bernabei *et al*. 2018) and a few false alarms, there is no *direct hint* about the nature of the dark matter, even if its *need* is widely acknowledged. We shall return to this point later in the text.

The case of dark energy is much more recent and even more difficult to expose, although it bears a deep similarity with the dark matter one in a broad sense: dark energy is invoked to explain a crucial dynamical observation (in this case, the accelerated expansion of the Universe) for which the "normal" and dark matter components give no solution at all. Let us sketch the reasoning leading to this hypothesis. A symbolic form of Einstein's theory of gravitation states that the space/time reacts to the presence of matter/energy (conceptually equated by Einsten's work), in turn creating paths for the latter to move, as if twisting a pipe carrying water. This is the *curvature* property of spacetime, and therefore we may verbally state Einstein's equations as

$$Variation\ of\ geometry\ =\ matter + energy\ content \quad (1)$$

when dealing with the Universe itself as the system to be described by the symbolic eq.(1), it is evident that *all* forms of matter and energy had to be counted in the right hand side. According to Einstein, these forms include not only energy densities, but also pressures, which is also a form of energy. Additional (metaphysical) hypothesis for the description of the Universe includes the isotropy and homogeneity of the cosmological solutions (i.e., no special places or directions), which may increase or decrease its spatial scale with cosmic time $t$. If we denote with $a(t)$ the so-called *scale factor* of the Universe, a concrete and explicit form of eq.(1), valid in the simplest case of spatially homogeneous and isotropic universes (first studied by Friedmann, Robertson and Walker) reads

$$\frac{\ddot{a}}{a} \propto -(\rho_{normal} + \rho_{DM} + \rho_{DE} + 3P_{normal} + 3P_{DM} + 3P_{DE}) \quad (2)$$

usually the "matter" components (normal and dark) do not contribute to the pressure, hence $P_{normal} = P_{DM} = 0$ is often assumed to simplify the picture. The root of the acceleration problem is seen on the left hand side: the $\ddot{a}$ is nothing but the second variation in time of the scale factor, that is, the analogous of acceleration in Newtonian physics. But if a positive acceleration must be obtained, as inferred from observations, then the "minus" sign in front of the right hand side must be compensated by a *negative* total of the quantities inside the parentheses. Since the normal and dark contributions to the pressure had already been neglected, this means that we must have $P_{DE} < -\frac{1}{3}(\rho_{normal} + \rho_{DM} + \rho_{DE})$, which leads to (remembering that energy densities can not be negative) to conclude that

$$P_{DE} < 0 \quad (3)$$

thus, if the Universe accelerates its expansion because of a new component (*dark energy*), its pressure must be negative, which is quite unusual for the kind of "fluid" approximation assumed in the cosmological setting.

Even though the hypothesis of a new component could be argued on the basis of methodological arguments, there is an additional feature that should be considered in favor of it: when an evaluation of the *fraction* of dark energy is performed, the obtained value is precisely the one astronomers had been seeking for a long time to complete several puzzles of cosmology (structure formation and the like), ~ 70% of the total balance required to have exactly a "closure" of the Universe (see, for example, Horvath 2009 for a discussion). Astronomers and physicists then talk about a *coincidence cosmology*, in which in spite of the total ignorance of the nature of the components, acknowledges that their relative contributions add up to a desired number and creates a consistent picture of the cosmic evolution.

The present situation is philosophically very disappointing (and even described sometimes as "embarassing", Rees 2003): instead of being able to identify and know the content of our Universe, we face a situation in which > 95% of it is unknown, the total fraction that belongs to the sum of dark matter and dark energy densities. As a corollary, the sometimes called "third Copernican revolution" ensues: we are *not* made of the same substance than the vast majority of Universe is made. Instead, we have to contemplate ourselves as a kind of residual stuff that managed nevertheless to arrange itself into structures we know as "human beings". In our opinion, this statement by itself is a major philosophical problem, to be added to the traditional list of fundamental questions on an equal footing. It is this cosmic perspective that we should have in mind when a deeper inquiry on the dark components is made. Note that a solution for this quandary may be attempted by modifying the "*Variation of geometry*" term on the left of eq.(1), without introducing new matter/energy components on the right hand side. We shall return to this possibility later.

## 3. Fundamental Building Blocks: timeline of the main ideas.

The classification of dark matter and dark energy must be seen in the context of a more ample inquiry, namely within the realm of the old quest for the fundamental building blocks of the world. About two millennia ago of philosophical investigations by Leucippus and Democritus on the Eleatic problem (Guthrie 1979) led these thinkers to the formulation of the *atomism*, in which discrete units roamed the "non-being" (in the sense of Parmenides) which served as an arena for their motion, the vacuum. Atomism was a quite different vision of the constitution of the physical world, carrying a strong materialistic content. The exposition by Lucretius in the 1st century B.C.E. was important to summarize the early atomists views, Epicurus work and his own contributions. Later on the atomism was "rediscovered" by Francis Bacon, Marin Mersenne, Robert Boyle and Isaac Newton (Meniel 1988, Kargon 2020, Stanford 2020d). It is the introduction of Newtonian space and universal time that gave a new twist to the old atomistic ideas, since space became a kind of recipient for matter (atoms) in which they move, collide and eventually formed all the existing variety of things.

In this context it can be said that our own view of the physical world, albeit refined by modern science to substitute the Greek concept of atoms by *elementary particles*, and with a complex *vacuum* as the arena for their motion, is a strong heritage of those ideas. The ultimate materialistic content of the contemporary theory is, however, still a matter of debate

(Feigl 1962, Russell 2009) in spite that in the original atomistic view this content was strongly anchored. Ordinary matter (protons and neutrons present in the atomic nucleus) found to hold point-like components inside (quarks) which show a peculiar phenomenological feature: in contrast to all other particles, they have never been seen as isolated entities, but only in pairs or triplets, the former corresponding to pions and similar light carriers and the latter to protons, neutrons and heavier particles. Since free quarks have not been detected isolatedly, there is some concern about their *existential dependence*. However, this holds just for the low-energy world: experiments have shown the production of a different state (Rafelski 2020) in which quarks are not confined into "bags" (hadrons) anymore. The role of the vacuum in which quarks live is thought to be crucial. As it stands, the quantum vacuum of modern physics is very different from the pre-20th century one, and quantum theory has indicated the presence of fluctuations that render the vacuum as a kind material medium, far from the "empty" previous classical picture. Hence we do not see any serious objection against quarks as fundamental building blocks of Nature, but remind that the quantum vacuum should enter somewhere in the fundamentality issue.

Meanwhile there is another evolving aspect of the view of the world that is important for our task, namely the relation of the physical world (existence) with the mathematical description of it. Pythagoras and the pythagoreans are probably the first thinkers to formulate the world as mathematics, the true path to its understanding according to the Crotone school. This question has been revisited and revived many times, in its modern form we can quote the position of E. Wigner who wrote about the "unreasonable effectiveness of mathematics" more than once (Wigner 1960). Einstein was actually inspired by Leibniz's relational theories, and his work was important for the idea of identifying physical existence with mathematical models, holding a realistic position he defended against the "perspectivism" of Quantum Theory, quite opposed in philosophical essence to his Relativity Theory approach. Mathematics as the generator and basis of the physical world is also present in other contemporary authors (Wolfram et al. 2020, Tegmark 2014).

The last issue in this cursory account is related to the controversy of 'energeticism' defended by W. Ostwald against L. Bolzmann (Deltete 2007, 2008), challenging the atoms in favor of energy as a more fundamental quantity. In spite of Boltzmann atoms being vindicated (i.e. Brownian motion explanation by Einstein), we may think that the fundamental issue of our work brings in some flavor of this issue, given that a loose component of the Universe (Dark Energy) is supposed to dominate its dynamics (see Section 8 below).

**4. Fundamental Building Blocks: their Mathematical Classification.**

As it is known, matter at the lowest compositeness level is presently identified with a set of *mathematical features,* which unlike the classical philosophical thought, these are totally unrelated to human senses. In other words, instead of the ancient philosophical classification into categories (i.e. hardness, shape, etc.) of primary, secondary or higher importance, we now talk about *mass, spin*, charge and other features which constitute the present classification of the physical entities. This shift was discussed by Burtt (2003) and many others.The introduction of mathematical entities in place of Aristotelian, Kantian and similar categories mainly based on human perception/logic is certainly a major step, although it was already latent in the heart of Greek atomism. The connection of such elementary mathematical properties defining the substance with the older macroscopic view of elementary matter of the pre-Newtonian thinkers is difficult and led to serious questioning

(see *Substance* in Stanford 2020b). But it is clear that there are strong reasons to argue forcefully that *all microphysical objects* measured so far fall into one of the particular classes which are classified by science by a general mathematical scheme. Therefore, we have a definite and rigorous framework to attempt a general classification of matter and energy (visible and "dark"), at least at the most fundamental level. Let us explain this point in some detail.

The relativistic spacetime, a fusion of spatial and temporal coordinates achieved at the turn of the century by Minkowski, Lorentz, Einstein and others, became the arena of all the events of the physical world. This means that there is a *group of transformations* from one frame to another that must be respected for any phenomenon occurring in spacetime. These transformations are *rotations, translations in space* and *translations in time* (the latter also known as *boosts* in the physics jargon). This set of transformations form a mathematical structure known as the *Poincare group*. It can be proved that with some of the mathematical 4-dimensional objects related to the Poincare group, two invariant quantities can be formed: the square of the 4-momentum $P^\mu$ and the square of the so-called Pauli-Lubanski vector $W^\mu$, which can be written as

$$P^\mu P_\mu = -m^2 \qquad (4)$$

$$W^\mu W_\mu = -m^2 s(s+1) \qquad (5)$$

where $m$ is a quantity identified with mass and $s$ is the value of the intrinsic angular momentum (spin) of the state (Wigner 1959, Ramond 1997)

Although the Poincare group is an abstract structure satisfying a set of rules, it can be *represented* on a variety of sets, called its *representations*. A few mathematical representations of the Poincare group are known, but the most interesting ones seem to have *positive discrete* mass values and *discrete* (integer or half-integer) spin numbers (for example, there exist imaginary mass-continuous spin properties but these do not seem to be physically realized in Nature, meaning that there is no hint of any detection of such states). Two particular cases of these representations seem to comprise all the known (empirically detected) particles: the bosons in the *adjoint* representation of integer spin, and the fermions in the *fundamental* representation with half-integer spin. Therefore, we may identify the "things" of the ontological categorization with the mass and spin numbers of the representation.

On the other hand, the elementary components also carry one or more "charges", which are introduced by the so-called direct product of an (internal) symmetry mathematical group times the spacetime Poincare´s group. Since these quantum numbers of generalized "charge" are actually not mixed with the spacetime transformations, acting in the "internal" space and serving to classify the states, it is not difficult to associate them with the "properties" of an ontological category. In fact, a recent work (Gilton 2020) discussed this difference among "mass" and "charge" pointing out their different ontological status and different mathematical descriptions as stated above. We point out that there have been attempts to unify both classes using the so-called *super-Poincare algebra*. However, a necessary consequence is that bosons and fermions are *not* separated anymore, rather they are united by supersymmetry. In spite of many efforts no hint of super symmetry has been observed in experiments, even for energies exceeding 1 Tera electron Volt in center of mass of colliding particles, but the quest for this feature continues with the analysis of data. Nevertheless, it is important to remember that "things" (mass and spin) and "properties" (charges) may not be different after all.

We can now turn our attention to the issue of the "dark" components of the Universe. As we shall see, the present ontological status of dark matter and dark energy is far from clear. In the next sections we shall sketch the possible classification of the former in a general framework anchored in the Poincare group representation as a benchmark concept.

**5. Do Dark Matter/Dark Energy really exist? Variations making them unnecessary**

We have mentioned above that an attempt to modify the left hand side terms of eq.(1) could be attempted. This possibility comprises a modification of gravity itself (i.e., *without* the requirement of dark components), which are currently under study, and it is pretended that, if successful, there is a hope that they would make the entire quest for the ontology of dark matter/energy *disappear*. That is, we would refine our view of how gravity works without introducing anything new in the microphysics world. Dark matter/energy would have a trivial ontological solution: they are not new entia at all, but rather features of the gravitational interaction, now refined beyond General Relativity (see the right side of Fig.1). One fundamental problem is that no general principle to guide the form/content of a new theory has been identified, and therefore the number of candidates is large, grouped with the name of "Alternative Theories of Gravity". By themselves, they may or may not need modifications of the ontological category.

There is also a large and heterogeneous class of models in which dark matter and dark energy are actually "geometric", in the sense of being related to higher dimensional effects. This class of models emerge by considering that the number of dimensions of the Universe is actually larger than four, and we actually live in a geometrical section of it (called the "brane", a generalized name for "membrane") whereas gravity propagated also into the extra dimensions (called the "bulk"). The projection of higher dimensional objects onto the brane leaves some "residuals" which are interpreted by we, 4-dimensional creatures, as the dark components. It is not clear whether these brane-world residuals must necessarily be quantized. For example, fermions and bosons on the brane have been studied, but it is not obvious that such entities must exactly respect the Poincare symmetries, although they may be constructed to do so. The popular phenomenological models known as *quintessence* (with a clear connection to the Medieval thought as an linguistic and conceptual homage) are constructed from their energy density and pressure, strongly suggesting that they actually are a way of integrating these dark components without an obvious connection to microphysics, at least at first sight, in spite that many of them can be related to microphysical models. The issue of dark matter/dark energy ontology would be then related to a very novel feature, namely an extra-dimensional Universe, and not necessarily rigidly tied to the Poincare group. But again, if this possibility is realized, it would constitute a very major advance in the physical and philosophical knowledge of our Universe by itself (right side of Fig. 1).

As an important general remark it must be said that the "old" problem of the existence of some dark matter component has a variety of evidences in support, from galactic to supercluster scales. From dynamical studies we may say that if DM exists it is *clustered* similarly to ordinary matter, and in fact it is quite difficult to get rid of it in all systems at once (Zavala et al. 2019). On the other hand, the evidence in favor of a *delocalized* DE is exclusively cosmological, and hence weaker because of this reason in the view of many members of the scientific community. Attempts to make DE superfluous have appeared in the literature (i.e. Lee 2020, Rácz et al. 207), although DM is also disputed from time to time (Mannheim, 2019). A very skeptical position is to consider DM/DE as epicycles (Bothun 2013).

## 6. Would we fit Dark Matter and Dark Energy within the Realm of Existing Matter/Energy Classes?

Provided DM/DE are not an effect of gravity, nor a projection of the brane world, and before embarking in a discussion about the different possibilities of the dark components, we must state that there is among scientists a "hidden", yet forceful assumption about the need of considering *quanta* of dark matter/energy. We shall see that in some cases a classical field would be enough, but the conviction that all matter is *quantized* drives the minds (and the language) towards the idea that some elementary particle (quantum), not yet identified, is a prime suspect for the dark component(s). This quantum nature of the whole world is a meta-postulate patterned after the success of Quantum Mechanics as a description of the physical world. By Occam's razor argument, it is very unlikely that the idea of a *classical* elementary component could be given serious consideration, and any such thing would be considered as a phenomenological provisional description at most.

We have mentioned that the present concept of (quantum) vacuum, or fundamental state is actually interesting and revealing: a long way has elapsed since the introduction of this concept by the Atomists to solve the Eleatic problem and giving a meaning to the Non-Being now identified with the vacuum.Thus, the atomistic vacuum became an *arena* for the atoms to move and collide, but meant an absolute empty. However, in the 20th century a redefinition of the vacuum concept according to quantum theory took place. The essential difference is that the vacuum is not, in its modern form, a state of "emptiness", but rather the state of the *lowest energy density*. In fact, quantum physics has shown that quantum fluctuations emerge and annihilate from the vacuum, leading to attribute a non-zero energy density to the vacuum. Moreover, this phenomenon has been tested in several cases, and led to definite predictions, later measured in laboratory (notably the Casimir effect, an attraction between two uncharged plates as a result of the quantum fluctuations existence, among others).

Since the Universe is essentially a vacuum to a large degree, even a rough calculation to estimate the cosmic energy density should render a sensible result. However, the numerical value thus obtained is so much larger than the one inferred that the identification cannot be blindly asserted. On the other hand, Zel'dovich (1968) was the first to argue that the transformation of the vacuum under the Poincare group operations leads to a *constant* term in Einstein's equations, and thus it should be identified with the term Einstein himself called his "greatest blunder", presently known as the *cosmological constant* $\Lambda$ (Horvath 2009). This numerical discrepancy between the inferred value of the cosmological constant $\leq 1$ and the theoretically calculated $\gg 1$ is still unsolved. However, if a reason for a numerical suppression of the value of $\Lambda$ is finally found, the simplest form of this delocalized dark energy will actually correspond to the simplest case, a trivially transformed quantity of the Poincare group, complying with our criterion. Of course, even if simple, this is far from being the only option for the dark energy nature.

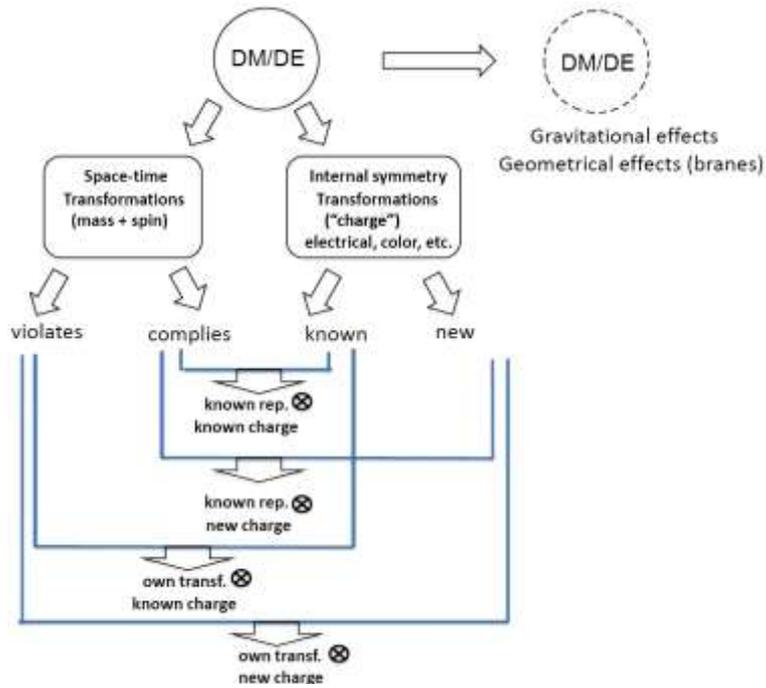

Fig. 1. A diagram representing the ontological possibilities of the Dark Matter and Dark Energy ("DM/DE"). The arrow to the right points out the possibility that they are "non-beings", either unknown effects of the gravitational field or "projections" of a higher-dimensional world (brane world) onto our 3+1 dimensional reality. The space-time transformations are characterized by mass and spin within the Poincare Group (or different labels within other possibilities, see section 6). The "Internal" symmetry classifies the object with a "charge" in a generalized sense. It must be remembered that a "new charge" can also be zero, but otherwise only symmetry principles could guide its assignation. The symbol "$\otimes$" represents the direct product of both the space-time and internal symmetry groups as defined in their mathematical formalism.

Even though the imagination of the researchers has been prodigious for creating models of dark matter/energy, it is important to see how are they linked to the Poincare group discussion above. Let us present some of the most popular choices and discuss how they relate to the ontological aspect, in the sense defined by the membership of a irreducible representation of the Poincare group times an internal symmetry.

The first and most straightforward possibility is that the dark components are actually members of a "dark sector", with couplings to ordinary matter which are very weak and thus remain "invisible" if not for their gravitational effects. In the literature there are examples of this class of theories constructed either respecting "extra" symmetries (i.e. ELKO and related proposals, see for example Ahluwalia & Nayak 2019). In many senses these relatives of known matter quanta can be considered as extensions that can be accommodated without a major harm to the Poincare scheme. The very popular supersymmetric candidates is another relevant example: the extension of the mathematical structure to the super-Poincare algebra would lead to supersymmetric partners of known particles which respect the Poincare scheme but now allowing the transformation of bosons into fermions and vice-versa. We would call these possibilities a "straightforward" extension of the Poincare group and its classification scheme, and would be covered by the central two merging arrows of Fig. 1 ("known representation" $\otimes$ "known charge" or ("known representation" $\otimes$ "unknown charge" if they carry a new quantum number to be discovered).

## 7. What if Poincare's Group is *not* a good description of the spacetime? The problem with mass and spin

Another, more involved possibility is that the transformations of the Poincare symmetry can be violated. Many models invoking, for example, violation of the Lorentz group (which is a subgroup of the Poincare group) have been presented (see for example, Arun, Gudennavar & Sivaram 2017 for a review). In this case we may have to consider alternative symmetries and mathematical groups. One important case is the substitution of the Poincare group by the de Sitter group, a natural choice if Nature happens to possess a *fundamental length* parameter. Since it can be shown that a fundamental parameter $\ell$ with the dimensions of length can be related to a constant term in Einstein´s equations (the cosmological constant $\Lambda$ mentioned before), then the "labels" of the mass and spin invariants should also contain a term related to the Cosmos itself, and although this is surely very small, conceptually it should change the way we look to an electron or a quark forever. As we see, changing the fundamental symmetry of spacetime has serious consequences for microphysics, but this bold possibility has been seriously considered (see for example, Aldrovandi, Beltran Almeida & Pereira 2004 for a consideration of a *de Sitter Relativity*). This would be achieved by changing the reference spacetime structure in the left square of Fig. 1, but would leave the rest untouched except for the "label" problem associated to the mass and spin of particles, pointing to an interdependence of the largest and smallest structures in our world.

## 8. Summary

We have adopted in the present work a variant of the *metaphysical foundationalism* position, insofar the mathematical structures understood as a fundamental element defining the objects of reality. Even though this does not imply a Pythagorean philosophy per se, the latter extreme attitude could be defended as well. All we are saying here is that the physical world is a sort of merge of physics and mathematics, in the sense that physics is based and finds support in mathematical structures, and thus is not "purely physical", at least from an empiricist's point of view. From that position we have outlined a classification of the status of the so-called "dark" components, which has a hierarchy of assumptions leading to their ultimate understanding. This ladder of this programme can be now explicitly formulated as

i) The dark components are necessary for the explanation of large-scale physics (galaxies, clusters, Hubble expansion and so on), and exist as separate entities. This is already questionable, as explained above, because a modified gravity scheme or higher-dimensional theory could be devised to provide a viable model, but is in fact a widely considered, dominating option.

ii) Provided i) is realized, there is a quantum explanation for both components, in the sense that some particle(s) not yet known need to be found to materialize the components nature. Non-quantum explanations are considered, at best, provisional, effective phenomenological descriptions at most.

iii) Given that i) and ii) are true, the dark quanta may or may not belong to the class of irreducible mathematical representations of the Poincare group, a mathematical scheme which includes all the known matter/energy quanta of the material Universe so far, times an "internal" symmetry group assignating some charge (which may be known or new, and even zero). If the fundamental symmetry of spacetime has to be changed (for example, turning to the de Sitter group outlined before or some other alternative), this will affect all the established knowledge in the elementary known world. A strange irony may arise if DM/DE quanta exist, and revive a form of the "energeticism" of W. Ostwald (Deltete 2007, 2008), since after a century of Boltzmannian matter, a form of energy would be proved a fundamental quantity in Physics.

We have tried to present a view of absolute fundamentality (Bennett 2017 ; Tahko, 2018 and references therein) for matter/energy and dark components, putting representations of a spacetime transformation group carrying generalized "charges" as independent entities, and sketch where the "dark" components could fit, provided they exist. If such a set can be constructed, they should form a complete minimal basis (CMB). But we have no fundamental objection against a future hypothetical discovery of a deeper level of fundamentality (preons, etc.), all it is assumed here is that the entia at hand are enough for the present level of known fundamental structure. A path towards a metaphysical infinitism (of the type implied by Bohmian type - Bohm, 1980-, or "wheels within wheels" structure without a bottom) or even a well-founded variant (Dehmelt 1989) is not closed, but it must be found by empirical research data and/or compelling theoretical constructions to proceed. Physicists have argued that the present level of structure (the Standard Model) has to be superseded, because it has no deep explanations for a number of problems, it features too many parameters and so on. The addition of DM/DE to this puzzle should be part of its solution in a broad sense including fundamental questions about the Universe.

It is presently not known what the final outcome of this quandary will be. However, it is certain to state that the situation is very disturbing, because statements that we ignore more than 95% of the content of our Universe are often expressed (Horvath 2009), as pointed out above. We may emphasize that we cannot conclude anything firm yet about the ontology of DM/DE, and some colleagues could add "or their very existence". We do need some breakthrough from the empirical side. The current steps taken to find out what are the dark matter and dark energy are twofold: an empirical avenue, intending to measure the so-called equation of state of the dark energy + experiments trying to detect the interaction of dark particles with known ordinary matter. However, we believe that these attempts are unlikely to provide *immediate* clues for the issue discussed here, because they make assumptions and seek evidence on the interaction only (i.e. gravitational effects only or cross-section for their scattering with matter), but do not directly probe their physical/mathematical nature in general. The quest for these elusive components of the Universe will continue for a long time to come (Zinkernagel 2002).